\documentclass[preprint,12pt]{elsarticle} 
\usepackage{fullpage}

\usepackage{amsmath}
\usepackage{amssymb}
\usepackage{bbm}

\usepackage{float} 
\usepackage{comment}
\usepackage{siunitx} 

\usepackage{tikzpagenodes}
\usepackage{tikz}
\usepackage{tikz-3dplot}
\usepackage{pgfplots}
\pgfplotsset{compat=1.18}
\usetikzlibrary{pgfplots.groupplots}
\usepackage{pgfplotstable}
\usetikzlibrary{calc}

\newcounter{bla}

\journal{the Journal of Computational Physics}

\usepackage{subcaption}
\usepackage{color}
\usepackage[colorlinks=true,linkcolor=blue, citecolor=blue, urlcolor=blue]{hyperref}
\usepackage[noabbrev]{cleveref}

\newcommand{\norm}[1]{\left\lVert #1 \right\rVert}

\begin{document}

\begin{frontmatter}



\title{SPIRAL: An Efficient Algorithm for the Integration of the Equation of Rotational Motion}



\author[a,b]{Carlos Andr\'es del Valle\corref{author}}
\author[a]{Vasileios Angelidakis}
\author[a]{Sudeshna Roy}
\author[b]{Jos\'e Daniel Mu\~noz} 
\author[a]{Thorsten P\"oschel}

\cortext[author] {Corresponding author.\\\textit{E-mail address:} cdelv@unal.edu.co}
\address[a]{
Institute for Multiscale Simulation, Friedrich-Alexander-Universit\"at Erlangen-N\"urnberg, Cauerstrasse 3, 91058 Erlangen, Germany.
}
\address[b]{   
Departamento de Física, Universidad Nacional de Colombia, Carrera 45 No. 26-85, Edificio Uriel Gutiérrez, Bogotá D.C., Colombia.
}

\begin{abstract}
We introduce \textsc{Spiral}, a third-order integration algorithm for the rotational motion of extended bodies. It requires only one force calculation per time step, does not require quaternion normalization at each time step, and can be formulated for both leapfrog and synchronous integration schemes, making it compatible with many particle simulation codes. The stability and precision of \textsc{Spiral} exceed those of state-of-the-art algorithms currently used in popular DEM codes such as \textsc{Yade}, \textsc{MercuryDPM}, \textsc{LIGGGHTS}, \text{PFC}, and more, at only slightly higher computational cost. Also, beyond DEM, we see potential applications in all numerical simulations that involve the 3D rotation of extended bodies.
\end{abstract}


\end{frontmatter}


\section{Introduction}
\label{sec:introduction}

The efficient and robust integration of the equations of rotational motion is a fundamental requirement for the simulation of many-particle systems. Developers of popular programs for the Discrete Element Method (DEM), see \autoref{tab:algorithms}, and Molecular Dynamics (MD), such as LAMMPS \cite{thompson2022lammps} and GROMACS \cite{abraham2015gromacs} face the problem of compromising between simple algorithms that require small time steps to maintain stability and more efficient algorithms such as Runge-Kutta-4 \cite{kutta1901}, that can handle larger time steps, but at the cost of a higher number of force calculations per time step, which can be prohibitively expensive for larger systems.
\begin{table}[htbp]
\centering
\caption{Popular algorithms for integrating the equation of rotational motion and DEM software that uses them. For a detailed discussion, see \ref{sec:Rotation_algorithms}.}
\label{tab:algorithms}
\begin{tabular}{l|l l l l}
\hline
algorithm & \multicolumn{3}{c}{DEM codes using it}\\ \hline
\footnotesize{Direct Euler} \cite{boyce2017}  & \footnotesize{\textsc{MFiX}}\cite{MFIX} & \footnotesize{\textsc{EDEM}}\cite{EDEM} & \footnotesize{\textsc{BlazeDEMGPU}}\cite{Blaze-DEMGPU} \!\!\!\!& \footnotesize{\textsc{MUSEN}}\cite{MUSEN}\\
\footnotesize{Velocity Verlet}\cite{Verlet}   & \footnotesize{\textsc{LIGGGHTS}}\cite{LIGGGHTS} & \footnotesize{\textsc{GranOO}}\cite{GranOO} & \footnotesize{\textsc{EDEM}}\cite{EDEM} & \footnotesize{\textsc{MercuryDPM}}\cite{MercuryDPM} \\
\footnotesize{Fincham} \cite{Fincham}         & \footnotesize{\textsc{Yade}}\cite{YADE} & \footnotesize{\textsc{Esys\_Particle}} \cite{EsysParticle} \!\!\!\!\!& \footnotesize{\textsc{WooDEM}}\cite{woodem} & \\
\footnotesize{Buss} \cite{buss2000accurate}   & \footnotesize{\textsc{PFC7}}\cite{PFC} & & & \\
\footnotesize{Johnson et al.}\cite{johnson2008quaternion} \!\!\!\!\!       & \footnotesize{\textsc{PFC7}}\cite{PFC} & & & \\
\footnotesize{PFC4} \cite{PFC4,ostanin2023rigid} & \footnotesize{\textsc{MercuryDPM}}\cite{MercuryDPM,ostanin2023rigid}\!\!\!\!\!\!\!\! & & & \\
         \hline
    \end{tabular}
\end{table}

When using quaternions to describe the particles' orientations in numerical algorithms, care must be taken to preserve their unitarity \cite{quaternions, quaternion_pdf}. Unfortunately, many popular algorithms do not conserve the quaternion norm, thus requiring normalization at each time step. \autoref{tab:algorithms} provides a list of algorithms utilized in widespread DEM codes to integrate the rotational motions of particles. Among those, Buss' algorithm \cite{buss2000accurate} stands out as it is the only one that does not require subsequent quaternion normalization. All algorithms listed in \autoref{tab:algorithms} provide second-order accuracy per time step, despite having better algorithms at our disposal for decades. For instance, Omelyan \cite{omelyan1998algorithm} proposed a third-order algorithm that shows remarkable energy conservation performance, a critical factor in MD simulations. Neto and Bellucci \cite{Neto_2006} proposed an algorithm that slightly outperforms Omelyan's in terms of energy conservation, albeit at the cost of higher complexity. 

We introduce an algorithm for integrating the equations of rotational motion of extended bodies that offers several advantages: It does not need quaternion normalization and requires only one force calculation per time step. It provides third-order accuracy and excellent numerical stability, including energy conservation, allowing for larger time steps and, thus, reducing overall computational costs. The algorithm can be adapted to different code architectures and simulation frameworks, including leapfrog and non-leapfrog variants, making it flexible and compatible with most simulation setups. 

In addition to mechanical many-body systems, numerical integration of the equation of rotational motion is used in many fields, including video game engines like Vulkan \cite{vulkan}, Unity \cite{unity}, and Unreal engine \cite{unrealengine}, animation and computer graphics \cite{kobilarov2009lie,johnson2003exploiting,grassia1998practical,su2009energy}, robotics \cite{todorov2012mujoco,beeson2015trac}, sensors \cite{sabatini2005quaternion,tadano2013three,valenti2015linear,sabatini2005quaternion}, navigation systems \cite{wu2005strapdown,chen2013modeling}, autonomous vehicles \cite{rucker2018integrating,pons2023quaternion,shah2018airsim}, and control systems \cite{tayebi2006attitude,manchester2016quaternion}. Remarkably, many control systems employ variational methods that result in a formulation similar to the one presented here, but they are much more complex \cite{rucker2018integrating,pons2023quaternion,sabatini2005quaternion} and generally provide only second-order accuracy. 

We compared the new integration algorithm against the previously mentioned ones, assessing accuracy, stability, performance, and energy conservation. We find excellent results across all the evaluation criteria, highlighting its potential as an improved general method for integrating equations for rotational motion.

\section{\textsc{Spiral}-an algorithm for the numerical integration of the equation of rotational motion}
\label{sec:AlgorithmDerivation}

The third-order \textsc{Spiral}\footnote{\textsc{Spiral} is an acronym for ``stable particle rotation integration algorithm''.} algorithm applies to leapfrog and non-leapfrog architectures. Here, we describe the leapfrog version. The non-leapfrog version is presented in \ref{app:Non-leapfrog}. We describe the 3D orientation of a rigid body at time $t$ by unit quaternions $q(t)$ \cite{Hamilton_quaternios}. To derive an expression for $q(t + \Delta t)$, we start with the quaternion's time derivative \cite{quaternions,quaternion_pdf},
\begin{equation}\label{eq:Dotq}
    \dot{q} = \frac{dq}{dt} = \frac{1}{2}\omega(t) q\,,
\end{equation}
where $\omega=\{0,\omega_x,\omega_y,\omega_z\}$ is a pure imaginary quaternion representing the angular velocity $\vec\omega=\{\omega_x,\omega_y,\omega_z\}$ in the reference frame of the 
rotating body. We assume that $\vec\omega$ depends only explicitly on time and solve the differential equation, \eqref{eq:Dotq} by separation of variables
\begin{equation}
    \int_t^{t + \Delta t} \frac{\text{d}q}{q} = \frac{1}{2}\int_{t}^{t + \Delta t} \omega(t^\prime) \text{d}t^\prime\,,
\end{equation}
to obtain 
\begin{equation}
\label{eq:TPq}
    q(t + \Delta t) = q(t)\,\exp\left(\frac{1}{2}\int_{t}^{t + \Delta t} \omega(t^\prime)\, \text{d}t^\prime\right) \,.
\end{equation}
We expand the unknown $\omega(t^\prime)$ in a Taylor series around $a$,
\begin{equation}\label{eq:TaylorOmega}
    \omega(t^\prime) = \omega(a) + \dot{\omega}(a)(t^\prime - a) +  \frac{1}{2}\ddot{\omega}(a)(t^\prime - a)^2 + \mathcal{O}\left[\left(t^\prime - a\right)^3\right]\,.
\end{equation}
Take $a=t + \frac{\Delta t}{2}$ (leapfrog assumption), and insert it into Eq. \eqref{eq:TPq}, to obtain
\begin{equation}
        q(t + \Delta t) = q(t)\,\exp\left[\frac{\Delta t}{2}\omega\left(t + \frac{\Delta t}{2}\right) + \mathcal{O}(\Delta t^3)\right]\,.
        \label{eq:leap}
\end{equation}
The exponential of a purely imaginary quaternion $\theta \hat{u}$, where $\theta$ is a scalar and $\hat{u}=\{0,u_x,u_y,u_z\}$, a unit quaternion, reads \cite{quaternion_pdf} 
\begin{equation}\label{eq:QuaternionPureRotation}
    e^{\theta\hat{u}} = \cos{\theta} + \vec{u}\sin{\theta}\,,
\end{equation}
with the unit vector $\vec u=\{u_x,u_y,u_z\}$. Then 
\begin{equation}
\label{eq:TPqfinal}
    q(t+\Delta t) = q(t)\left(\cos{\theta_1} + \frac{\vec{\omega}\left(t + \frac{\Delta t}{2}\right)}{\left|\vec{\omega}\left(t + \frac{\Delta t}{2}\right)\right|}\sin{\theta_1} \right),\quad \text{with} \quad
    \theta_1 \equiv \frac{\Delta t}{2}\left|\vec{\omega}\left(t + \frac{\Delta t}{2}\right)\right|\,.
\end{equation}
The shown transformations preserve the norm of the initial quaternion, thus eliminating the need for subsequent normalization. The approach is similar to the one by Buss \cite{buss2000accurate}, but Eq. \ref{eq:TPqfinal} achieves a higher-order precision. Also, Seelen et al. \cite{Seelen2016} propose a predictor-corrector version of the Zhao and Van Wachem scheme \cite{Zhao2013} by using a similar formulation, but achieving only second-order accuracy as well.

The expression for $q(t+\Delta t)$ is independent of the method chosen to compute $\vec \omega$. Nevertheless, for optimal results, accurate angular velocity calculation is crucial and non-trivial given the highly nonlinear expressions for angular velocity. We employ a Strong-Stability Preserving Runge-Kutta-3 (SSPRK3) scheme \cite{SSPRK3}, which is highly stable and requires fewer operations than Runge-Kutta-4 \cite{kutta1901} and the iterative algorithm used by Omelyan \cite{omelyan1998algorithm}. Moreover, inspired by \cite{johnson2008quaternion}, we compute the torque $\vec M(t)$ only once per time step, assuming the torque constant during the time step. Using SSPRK3, we update the angular velocity through
\begin{equation}\label{eq:algorithm_w}
\vec{\omega}\left(t + \frac{\Delta t}{2}\right) = \vec{\omega}\left(t-\frac{\Delta t}{2}\right) + \frac{1}{6}\left(\vec{K}_1 + \vec{K}_2 + 4\,\vec{K}_3\right) \,,
\end{equation}
with 
\begin{equation}
\begin{split}
\vec{K}_1 &= \Delta t\,\dot{\vec{\omega}}\left(\vec{\omega}, \vec{M}(t)\right) \, ,\\
\vec{K}_2 &= \Delta t\,\dot{\vec{\omega}}\left(\vec{\omega} +\vec{K}_1, \vec{M}(t)\right) \, ,\\
\vec{K}_3 &= \Delta t\,\dot{\vec{\omega}}\left(\vec{\omega} + \frac{1}{4}\left(\vec{K}_1 + \vec{K}_2\right), \vec{M}(t) \right) \, ,
\end{split}
\end{equation}
where $\vec{M}(t)$ is the torque acting in the body's principal axis reference frame. The acceleration $\dot{\vec \omega} = \left\{\dot{\omega}_x, \dot{\omega}_y, \dot{\omega}_z\right\}$ is given by Euler's equations of motion \cite{Goldstein},
\begin{equation}\label{eq:ang_vel_derivative}
\begin{split}
\dot{\omega}_x &= \frac{M_x}{I_x}+ \omega_y\,\omega_z \frac{I_y - I_z}{I_x}\, , \\
\dot{\omega}_y &= \frac{M_y}{I_y}+ \omega_z\,\omega_x \frac{I_z - I_x}{I_y}\, , \\
\dot{\omega}_z &= \frac{M_z}{I_z}+ \omega_x\,\omega_y \frac{I_x - I_y}{I_z}\, ,
\end{split}
\end{equation}
with the body's principal moments of inertia $I_x$, $I_y$, and $I_z$. 
This SSPRK3 integration scheme for the angular velocity is an excellent choice for non-spherical particles; nevertheless, for spherical ones, we suggest using a normal leapfrog \cite{leapfrog} as the non-linearities vanish. Incurring in performance improvements without losing precision.  

Using the leapfrog scheme, $\vec \omega$ is updated before $q$. Therefore, it is necessary to compute the angular velocity back by half a step before starting the simulation. This can be done with Eq. \eqref{eq:algorithm_w} with the argument $-\frac{\Delta t}{2}$ instead of $\Delta t$. 

\section{Validation against an analytical solution}
\label{sec:analytical_sol}

Consider a rigid body with principal moments of inertia $I_x, I_y, I_z$. For the special case $I_x \neq I_y = I_z$ we can solve the set of nonlinear differential equations, Eq. \eqref{eq:ang_vel_derivative} analytically, provided a constant torque $\vec{M} = \{M_x, 0, 0\}$ with $M_x \neq 0$ and $\omega_x(t) \neq 0$ at all times. The solution reads (see \ref{app:AnalyticalSolution} for the derivation) 
\begin{equation}\label{eq:analytical_solution}
\begin{split}
\omega_x(t) & = \sqrt{\tau} = \omega_x^0 + \frac{M_x}{I_x}t, \\
\omega_y(\tau) &= k_1 \cos \left(\Omega\tau\right) + k_2 \sin\left(\Omega\tau\right)\,, \\
\omega_z(\tau) &= \eta\Omega \left(k_2 \cos\left(\Omega\tau\right) - k_1 \sin\left(\Omega\tau\right)\right)\,,
\end{split}
\end{equation}
with the initial angular velocity $\vec{\omega}(0) = \{\omega_x^0, \omega_y^0, \omega_z^0\}$ and
\begin{equation}
\begin{split}
\Omega^2 &\equiv -\left(\frac{I_x}{2 M_x}\right)^2 \frac{(I_x - I_y)(I_z - I_x)}{I_zI_y}\, ,\\
\eta &\equiv \frac{I_y}{I_z-I_x} \,\frac{2\,M_x}{I_x}\,, \\
k_{1} &\equiv \omega_{y}^0 \cos\left(\Omega \right(\omega_x^0\left)^2\right) - \frac{\omega_{z}^0}{\Omega \eta}\sin\left(\Omega \right(\omega_x^0\left)^2\right),\\
k_{2} &\equiv \omega_{y}^0 \sin\left(\Omega \right(\omega_x^0\left)^2\right) + \frac{\omega_{z}^0}{\Omega \eta}\cos\left(\Omega \right(\omega_x^0\left)^2\right).
\end{split}
\end{equation}
Note that this solution does not apply for $M_x = 0$ (for this case, the solution is also given in \ref{app:AnalyticalSolution}). Further analytically solvable special cases of the rigid body equation can be found in \cite{Analytic1,Analytic2,Analytic3}.

For subsequent comparison, we use the above analytical expression of the body's angular velocity (which we call $\vec \omega'$) to compute its orientation $q^\prime(t)$ by numerical integration of \eqref{eq:Dotq} using the highly accurate \textit{DifferentialEquations.jl} \cite{Differentialequations.jl}. The orientation $q^\prime(t)$ will serve as a reference to quantify the accuracy of $q(t)$ as delivered by our algorithm compared to other algorithms from the literature.

To compute the error between the analytical solution and the solution from the different integration algorithms, we define the error metric:
\begin{equation}\label{eq:error_metric}
    \norm{\vec{v} - \vec{v}^\prime} := \frac{\sum_i |v_i - v^\prime_i|}{\sum_i |v^\prime_i|}.
\end{equation}
This metric is the vector's relative $L_1$ norm. We use the $L_1$ norm instead of $L_2$  because $L_2$ measures the drift of the norm and the error in the direction; in contrast, the $L_1$ norm gives more weight to the error in individual components of the vector. 

\begin{figure}[htbp]
    \centering
    \begin{subfigure}[b]{0.47\textwidth}
        \centering
        \begin{tikzpicture}
            \begin{axis}[
                xlabel=time {[\si{s}]},
                ylabel=$\omega_i^\prime$ {[\si{rad/s}]},
                legend style={at={(0.5,1.05)}, anchor=south, legend cell align=left, legend columns=-1, column sep=0.2cm},
                width=\textwidth,
                no markers,
                every axis plot/.append style={very thick},
                ymax = 12,
                xmax = 1.2,
            ]
                \addplot[blue, densely dotted] table [x=Rkt_Time, y=Rkt_w1, col sep=comma] {analytical_sol.csv}; 
                \addlegendentry{$\omega_x^\prime$}
                \addplot[red, densely dashed] table [x=Rkt_Time, y=Rkt_w2, col sep=comma] {analytical_sol.csv};
                \addlegendentry{$\omega_y^\prime$}
                \addplot[black, densely dash dot dot] table [x=Rkt_Time, y=Rkt_w3, col sep=comma] {analytical_sol.csv};
                \addlegendentry{$\omega_z^\prime$}
            \end{axis}
            \rlap{\raisebox{0.57\dimexpr\height}{\makebox[0.77\textwidth][r]{
                \includegraphics[width=0.55\textwidth]{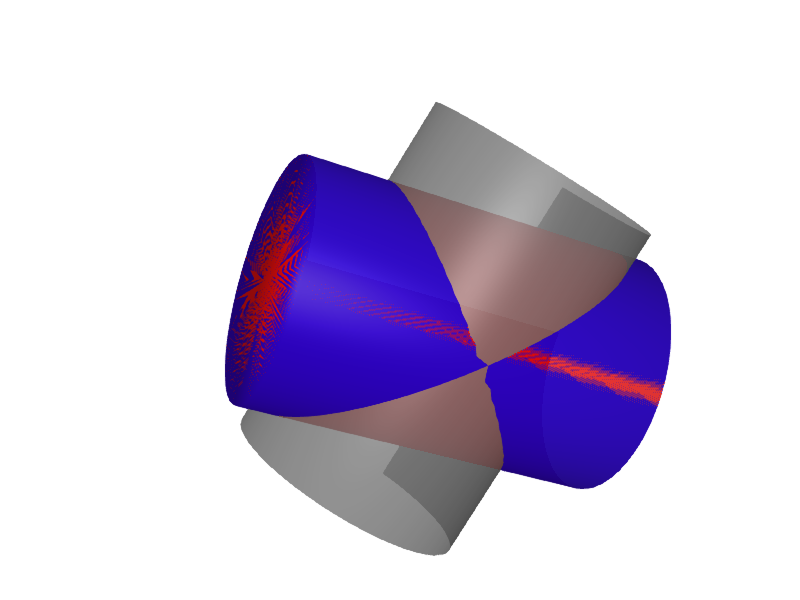}
            }}}
            \node[anchor=south west] at (0.32\textwidth,1.5) {\scriptsize{orientation at $t=15\,\text{s}$}};
        \end{tikzpicture}
        \caption{components of the angular velocity}
        \label{subfig:omegaprime}
    \end{subfigure}
    \begin{subfigure}[b]{0.47\textwidth}
        \centering
        \begin{tikzpicture}
            \begin{axis}[
                xlabel=time {[\si{s}]},
                ylabel=$q_i^\prime$,
                legend style={at={(0.5,1.05)}, anchor=south, legend cell align=left, legend columns=-1, column sep=0.2cm},
                width=\textwidth,
                no markers,
                every axis plot/.append style={very thick},
                xmax = 1.2,
            ]
                \addplot[blue, densely dashed] table [x=Rkt_Time, y=Rkt_q1, col sep=comma] {analytical_sol.csv}; 
                \addlegendentry{$q_1^\prime$}
                \addplot[red, densely dotted] table [x=Rkt_Time, y=Rkt_q2, col sep=comma] {analytical_sol.csv};
                \addlegendentry{$q_2^\prime$}
                \addplot[brown, densely dash dot dot] table [x=Rkt_Time, y=Rkt_q3, col sep=comma] {analytical_sol.csv};
                \addlegendentry{$q_3^\prime$}
                \addplot table [x=Rkt_Time, y=Rkt_q4, col sep=comma] {analytical_sol.csv};
                \addlegendentry{$q_4^\prime$}
            \end{axis}
        \end{tikzpicture}
        \caption{quaternion components}
        \label{subfig:qprime}
    \end{subfigure}
    \caption{3D rotation of a cylinder subjected to a constant torque: (\subref{subfig:omegaprime}) analytical solution for the angular velocity $\vec{\omega}^\prime$, (\subref{subfig:qprime}) highly accurate numerical reference solution for the orientation $q^\prime(t)$. For the detailed set of parameters, see the text. The inset to (\subref{subfig:omegaprime}) shows the cylinder's orientation at $t = 15\;\text{s}$ for the reference solution (red), our proposed \textsc{Spiral} algorithm (blue, nearly coinciding with red), and the direct Euler algorithm (grey).}
    \label{fig:analytical_sol_plot}
\end{figure}

The analytic solution, Eq. \eqref{eq:analytical_solution}, describes an object with two identical principal moments of inertia, e.g., a cylinder, subjected to a constant torque. Here, we consider a steel cylinder of radius 5\,\si{cm}, height 15\,\si{cm}, and density 7750\,\si{kg/m^3}. Then, the corresponding tensor of inertia in the cylinder's principal axis frame is $\Vec{I} = \{I_x,I_y,I_z\}=\{0.0114,0.0228,0.0228\}\,\si{kg.m^2}$. We further assume the cylinder's initial angular velocity, $\vec{\omega_0} = \{0.3, -0.9, 0.6\}\,\si{rad/s}$, its initial orientation $q_0 = \{1, 0, 0, 0\}$, and the time-independend torque, $M_x = 0.5\,\si{N.m}$. Fig. \ref{fig:analytical_sol_plot} shows the analytical solution for the specified parameters.

\begin{figure}[htbp]
    \centering
    \resizebox{\textwidth}{!}{%
    \begin{tikzpicture}
        \begin{groupplot}[
            group style={
                group size=2 by 1, 
                horizontal sep=2.4cm, 
            },
            axis x line=bottom,
            axis y line=left,
            xlabel=$\Delta t$  {[\si{s}]}, 
            no markers,
            every axis plot/.append style={very thick},
            xmode=log,
            ymode=log,
            yminorticks=false,
            xminorticks=false
        ]
        
        \nextgroupplot[ylabel=relative error:~ $\norm{q - q^\prime}$]
        \addplot [color=orange, loosely dashed] table [x=dt, y=Eulerq, col sep=comma] {Latex_dt_error.csv}; \label{plot:Eulerq}
        \addplot [color=yellow!80!black, loosely dotted] table [x=dt, y=MercuryDPMw, col sep=comma] {Latex_dt_error.csv}; \label{plot:Mercuryq}
        \addplot [color=green, loosely dotted] table [x=dt, y=Verletq, col sep=comma] {Latex_dt_error.csv}; \label{plot:Verletq}
        \addplot [color=black, densely dash dot] table [x=dt, y=Bussq, col sep=comma] {Latex_dt_error.csv}; \label{plot:Bussq}
        \addplot [color=cyan, loosely dash dot dot] table [x=dt, y=Johnsonq, col sep=comma] {Latex_dt_error.csv}; \label{plot:Johnsonq}
        \addplot [color=magenta, loosely dash dot] table [x=dt, y=Finchamq, col sep=comma] {Latex_dt_error.csv}; \label{plot:Finchamq}
        \addplot [color=blue, densely dashed] table [x=dt, y=Omelyanq, col sep=comma] {Latex_dt_error.csv}; \label{plot:Omelyanq}
        \addplot [color=red] table [x=dt, y=SPIRALLq, col sep=comma] {Latex_dt_error.csv}; \label{plot:SPIRALLq}
        \addplot [color=gray, densely dotted] table [x=dt, y=RKq, col sep=comma] {Latex_dt_error.csv}; \label{plot:RKq}
        
        \nextgroupplot[ylabel=relative error:~ $\norm{\vec{\omega} - \vec{\omega^\prime}}$]
        \addplot [color=orange, loosely dashed] table [x=dt, y=Eulerw, col sep=comma] {Latex_dt_error.csv}; \label{plot:Eulerw}
        \addplot [color=yellow!80!black, loosely dotted] table [x=dt, y=MercuryDPMw, col sep=comma] {Latex_dt_error.csv}; \label{plot:Mercuryw}
        \addplot [color=green, loosely dotted] table [x=dt, y=Verletw, col sep=comma] {Latex_dt_error.csv}; \label{plot:Verletw}
        \addplot [color=black, densely dash dot] table [x=dt, y=Bussw, col sep=comma] {Latex_dt_error.csv}; \label{plot:Bussw}
        \addplot [color=cyan, loosely dash dot dot] table [x=dt, y=Johnsonw, col sep=comma] {Latex_dt_error.csv}; \label{plot:Johnsonw}
        \addplot [color=magenta, loosely dash dot] table [x=dt, y=Finchamw, col sep=comma] {Latex_dt_error.csv}; \label{plot:Finchamw}
        \addplot [color=blue, densely dashed] table [x=dt, y=Omelyanw, col sep=comma] {Latex_dt_error.csv}; \label{plot:Omelyanw}
        \addplot [color=red] table [x=dt, y=SPIRALLw, col sep=comma] {Latex_dt_error.csv}; \label{plot:SPIRALLw}
        \addplot [color=gray, densely dotted] table [x=dt, y=RKw, col sep=comma] {Latex_dt_error.csv}; \label{plot:RKw}
        \end{groupplot}
        
        \node[align=center,anchor=north] at ($(group c1r1.south)!0.5!(group c2r1.south) + (0,-1.2cm)$) {%
         \begin{tabular}{@{}llll@{}}
                \ref{plot:RKq} Runge Kutta 4 \cite{kutta1901} & \ref{plot:Bussq} Buss \cite{buss2000accurate} & \ref{plot:Finchamq} Fincham \cite{Fincham} & \ref{plot:Omelyanq} Omelyan \cite{omelyan1998algorithm}\\
                \ref{plot:Eulerq} Direct Euler \cite{boyce2017} & \ref{plot:Johnsonq} Johnson et al. \cite{johnson2008quaternion} & \ref{plot:Verletq} Velocity Verlet \cite{Verlet} & \ref{plot:Mercuryq} PFC4 \cite{PFC4}\\
                \ref{plot:SPIRALLq} \textsc{Spiral} (this paper)\\
        \end{tabular}};
    \end{tikzpicture}}
    \caption{Comparison of various algorithms with the analytical solution at time $t=1\,\si{s}$: (left) relative error of the predicted quaternion with respect to the semi-analytical solution as a function of the time step, (right) relative error of the angular velocity with respect to the analytical solution as a function of the time step. The errors were computed using Eq. \eqref{eq:error_metric}.
}
\label{fig:dt_error}
\end{figure}
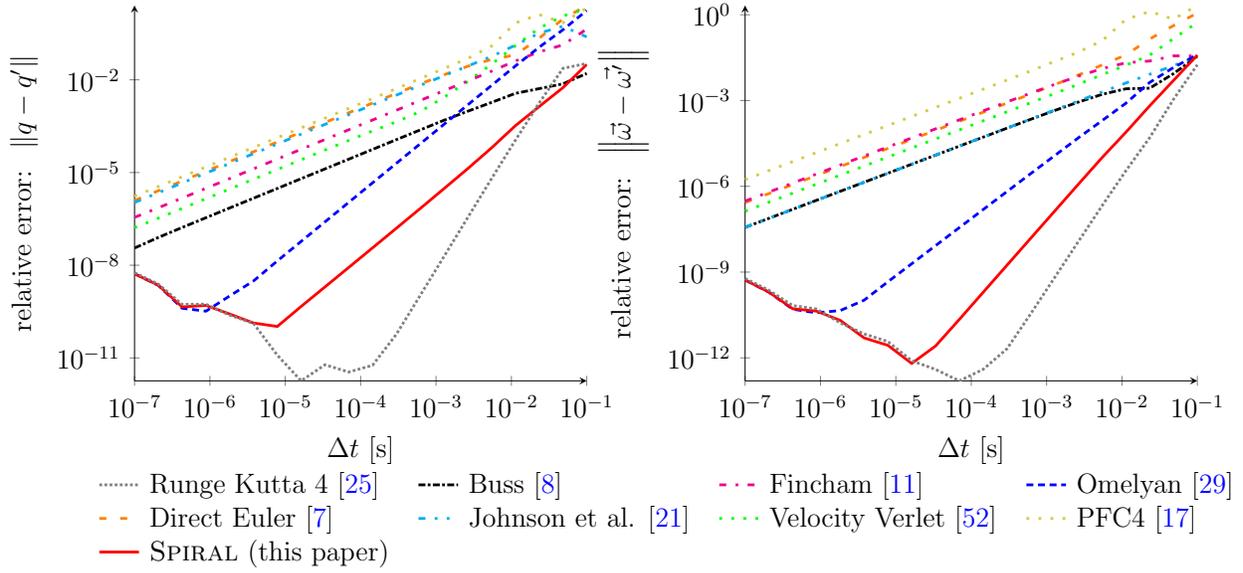

The precision of various integration algorithms is depicted in Fig. \ref{fig:dt_error}. It shows the error as defined in Eq. \eqref{eq:error_metric} at $t=1\,\text{s}$ between the numerically obtained values, $q$ and $\vec{\omega}$, and the (semi-)analytical results, $q^\prime$ and $\vec{\omega}^\prime$, respectively, as functions of the time step $\Delta t$.
We see that \textsc{Spiral} outperforms all other integration schemes studied over the entire range of time steps, some of them by several orders of magnitude. Only the Runge-Kutta-4-scheme \cite{kutta1901} performs better; however, this algorithm requires four torque computations per time step compared to one for \textsc{Spiral} and, therefore, Runge-Kutta-4 is not suitable for large-scale simulations. More details on the compared algorithms can be found in \ref{sec:Rotation_algorithms}. 

\begin{figure}[htbp]
    \centering
    \resizebox{\textwidth}{!}{%
    \begin{tikzpicture}
        \begin{groupplot}[
            group style={
                group size=2 by 1, 
                horizontal sep=2.4cm, 
            },
            axis x line=bottom,
            axis y line=left,
            xlabel=total simulation time $t$ {[\si{s}]}, 
            no markers,
            every axis plot/.append style={very thick},
            ymode=log,
            yminorticks=false,
            xminorticks=false
        ]
        
        \nextgroupplot[ylabel=relative error: $\norm{q - q^\prime}$]
        \addplot [color=orange, loosely dashed] table [x=Euler_Time, y=Euler_Error_q, col sep=comma] {qw_error_time.csv}; \label{plot:Eulerqt}
        \addplot [color=yellow!80!black, loosely dotted] table [x=Mercury_Time, y=Mercury_Error_q, col sep=comma] {qw_error_time.csv}; \label{plot:Mercuryqt}
        \addplot [color=green, loosely dotted] table [x=Verlet_Time, y=Verlet_Error_q, col sep=comma] {qw_error_time.csv}; \label{plot:Verletqt}
        \addplot [color=black, densely dash dot] table [x=Buss_Time, y=Buss_Error_q, col sep=comma] {qw_error_time.csv}; \label{plot:Bussqt}
        \addplot [color=cyan, loosely dash dot dot] table [x=Johnson_Time, y=Johnson_Error_q, col sep=comma] {qw_error_time.csv}; \label{plot:Johnsonqt}
        \addplot [color=magenta, loosely dash dot] table [x=Fincham_Time, y=Fincham_Error_q, col sep=comma] {qw_error_time.csv}; \label{plot:Finchamqt}
        \addplot [color=blue, densely dashed] table [x=Omelyan_Time, y=Omelyan_Error_q, col sep=comma] {qw_error_time.csv}; \label{plot:Omelyanqt}
        \addplot [color=red] table [x=SPIRAL_L_Time, y=SPIRAL_L_Error_q, col sep=comma] {qw_error_time.csv}; \label{plot:SPIRALLqt}
        \addplot [color=gray, densely dotted] table [x=Rkt4_Time, y=Rkt4_Error_q, col sep=comma] {qw_error_time.csv}; \label{plot:RKqt}

        \nextgroupplot[ylabel=relative error: $\norm{\vec{\omega} - \vec{\omega^\prime}}$]
        \addplot [color=orange, loosely dashed] table [x=Euler_Time, y=Euler_Error_w, col sep=comma] {qw_error_time.csv}; \label{plot:Eulerwt}
        \addplot [color=yellow!80!black, loosely dotted] table [x=Mercury_Time, y=Mercury_Error_w, col sep=comma] {qw_error_time.csv}; \label{plot:Mercurywt}
        \addplot [color=green, loosely dotted] table [x=Verlet_Time, y=Verlet_Error_w, col sep=comma] {qw_error_time.csv}; \label{plot:Verletwt}
        \addplot [color=black, densely dash dot] table [x=Buss_Time, y=Buss_Error_w, col sep=comma] {qw_error_time.csv}; \label{plot:Busswt}
        \addplot [color=cyan, loosely dash dot dot] table [x=Johnson_Time, y=Johnson_Error_w, col sep=comma] {qw_error_time.csv}; \label{plot:Johnsonwt}
        \addplot [color=magenta, loosely dash dot] table [x=Fincham_Time, y=Fincham_Error_w, col sep=comma] {qw_error_time.csv}; \label{plot:Finchamwt}
        \addplot [color=blue, densely dashed] table [x=Omelyan_Time, y=Omelyan_Error_w, col sep=comma] {qw_error_time.csv}; \label{plot:Omelyanwt}
        \addplot [color=red] table [x=SPIRAL_L_Time, y=SPIRAL_L_Error_w, col sep=comma] {qw_error_time.csv}; \label{plot:SPIRALLwt}
        \addplot [color=gray, densely dotted] table [x=Rkt4_Time, y=Rkt4_Error_w, col sep=comma] {qw_error_time.csv}; \label{plot:RKwt}
        \end{groupplot}
        
        \node[align=center,anchor=north] at ($(group c1r1.south)!0.5!(group c2r1.south) + (0,-1.2cm)$) {%
         \begin{tabular}{@{}llll@{}}
                \ref{plot:RKq} Runge Kutta 4 \cite{kutta1901} & \ref{plot:Bussq} Buss \cite{buss2000accurate} & \ref{plot:Finchamq} Fincham \cite{Fincham} & \ref{plot:Omelyanq} Omelyan \cite{omelyan1998algorithm}\\
                \ref{plot:Eulerq} Direct Euler \cite{boyce2017} & \ref{plot:Johnsonq} Johnson et al. \cite{johnson2008quaternion} & \ref{plot:Verletq} Velocity Verlet \cite{Verlet} & \ref{plot:Mercuryq} PFC4 \cite{PFC4}\\
                \ref{plot:SPIRALLq} \textsc{Spiral} (this paper)\\
        \end{tabular}};
    \end{tikzpicture}}
    \caption{Comparison of various algorithms with the analytical solution; evolution of the error. The time step is $\Delta t = 10^{-5}$\;\si{s}: (left) relative error of the predicted quaternion with respect to the semi-analytical solution as a function of total simulation time, (right) relative error of the angular velocity with respect to the analytical solution as a function of the time step. The error was computed according to Eq. \eqref{eq:error_metric}.}
    \label{fig:qw_erorr_time}
\end{figure}
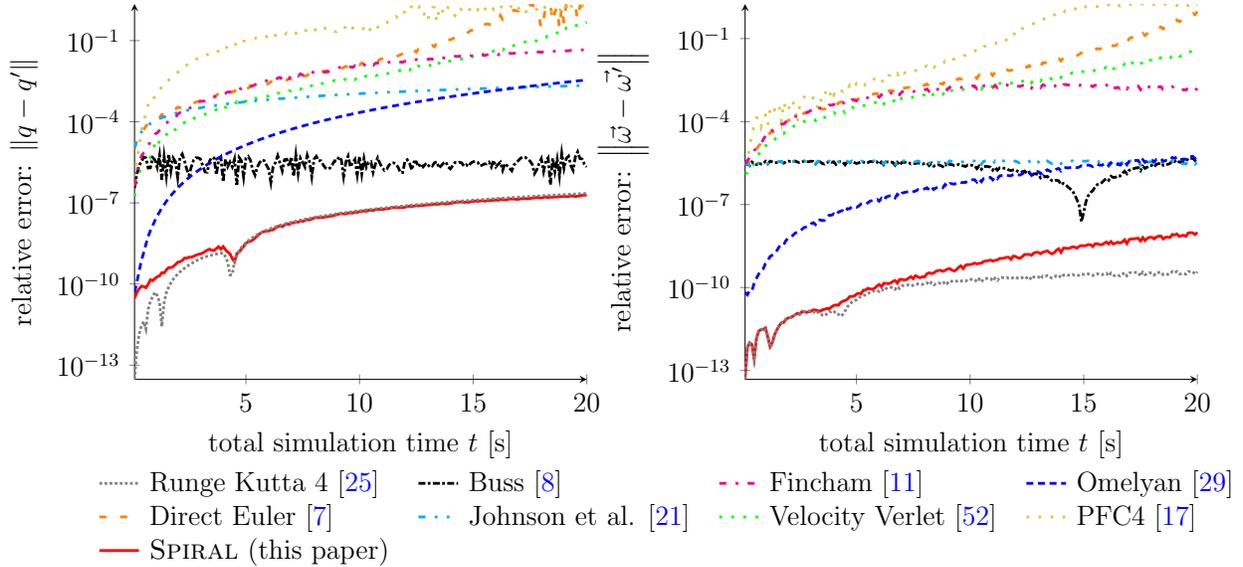

While Fig. \ref{fig:dt_error} provides valuable insights into the accuracy of the integration algorithms, it does not explicitly demonstrate their numerical stability. To assess stability, Fig. \ref{fig:qw_erorr_time} shows the accumulated relative error as a function of the total simulation time. The time step is chosen $\Delta t = 10^{-5}$\,\si{s}. 

Again, we see that \textsc{Spiral} outperforms the other algorithms, except for Runge-Kutta-4 \cite{kutta1901}, which uses four torque calculations per time step. In contrast to \textsc{Spiral}, all other integration algorithms in the plot require quaternion normalization after each time step, except for the scheme by Buss \cite{buss2000accurate}. We have noticed that the Omelyan algorithm \cite{omelyan1998algorithm}, albeit not requiring quaternion normalization due to its mathematical formulation, tends to accumulate significant numerical errors over time. Therefore, periodic normalization of the quaternions is advised.

\clearpage

\section{Performance analysis}
\label{sec:Performance}
\begin{figure}[H]
\centering
\resizebox{0.65\textwidth}{!}{%
\begin{tikzpicture}
\begin{axis}[
    bar width = 9pt,
    xbar=0,
    xlabel = execution time (normalized by Direct Euler \cite{boyce2017}),
    y=0.8cm,
    yticklabels = {\textsc{Spiral} (this paper), Buss \cite{buss2000accurate}, Direct Euler \cite{boyce2017}, Fincham \cite{Fincham}, Johnson et al. \cite{johnson2008quaternion}, PFC4 \cite{PFC4}, Omelyan \cite{omelyan1998algorithm}, Velocity Verlet \cite{Verlet}, Runge-Kutta 4 \cite{kutta1901}}, 
    ytick = data,
    yticklabel style={align=center},
    axis y line*=none, 
    legend style={at={(0.5,1.05)}, anchor=south, legend cell align=left, legend columns=2, column sep=0.2cm},
    legend image code/.code={\draw [#1] (0cm,-0.1cm) rectangle (0.2cm,0.25cm); },
]
\addplot[fill=blue!60!white, draw=black, error bars/.cd, x dir = both, x explicit] coordinates {
    (0, 0) +- (0.0, 0) 
    (0, 1) +- (0.0, 1) 
    (1.0, 2) +- (0.01889831972214469, 2) 
    (1.1061955533994414, 3) +- (0.15945347150770608, 3) 
    (2.763917462266236, 4) +- (0.11361994137522263, 4) 
    (1.9598387307338385, 5) +- (0.11762261299518967, 5) 
    (1.7418379418038126, 6) +- (0.2928766666180153, 6) 
    (0.9716795328093064, 7) +- (0.047909575668675514, 7) 
};
\addplot[fill=orange!70!white, draw=black, error bars/.cd, x dir = both, x explicit] coordinates {
    (1.335396119224087, 0) +- (0.04426661483177273, 0) 
    (1.8808185868129341, 1) +- (0.13546363971860714, 1) 
    (0.824897640931762, 2) +- (0.05826502501331162, 2) 
    (0.6968598154668283, 3) +- (0.03885278383958818, 3) 
    (2.607415748788213, 4) +- (0.16588269611404619, 4) 
    (1.2194868637799356, 5) +- (0.11640244619922989, 5) 
    (1.2317782094633314, 6) +- (0.030695781330393523, 6) 
    (0.7637010897918357, 7) +- (0.025018097129532332, 7) 
};
\legend{normalized, not normalized};
\end{axis}
\end{tikzpicture}}
\caption{Execution time per time step relative to the time required by the Direct Euler algorithm \cite{boyce2017}. Results with and without quaternion normalization. Note that \textsc{Spiral} does not require quaternion normalization.}
\label{fig:performance2}
\end{figure}
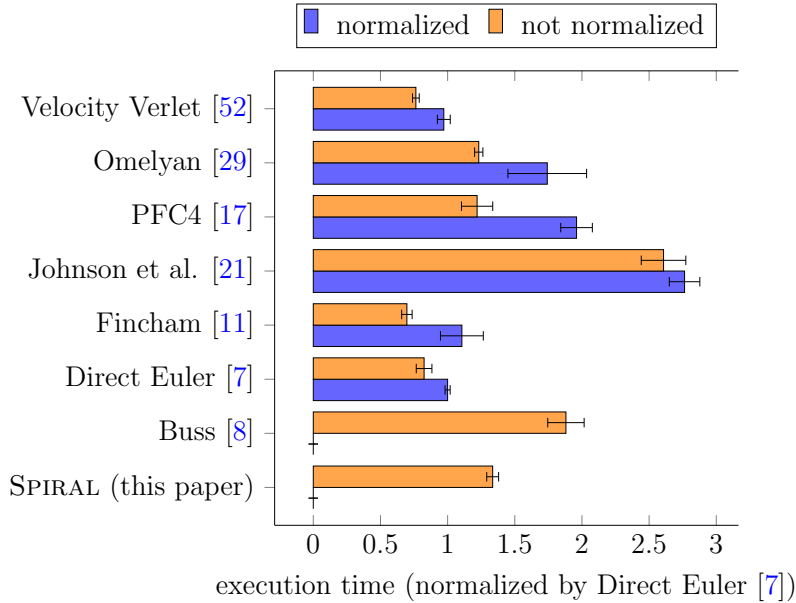

Fig. \ref{fig:performance2} shows the computer time of the algorithms for one time step, relative to the time employed by the Direct Euler algorithm when normalizing the quaternion \cite{boyce2017}.

The performance metrics were determined using \textsc{BenchmarkTools.jl} \cite{BenchmarkTools.jl}. We see that quaternion normalization involves significant performance degradation. Therefore, it is a clear advantage that \textsc{Spiral} does not require normalization.

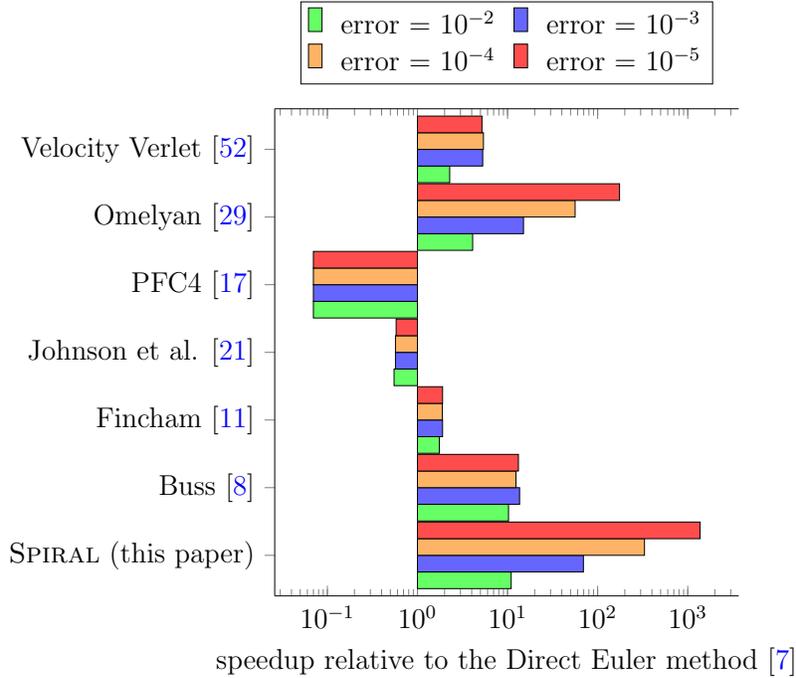
\begin{figure}[htbp]
\centering
\resizebox{0.65\textwidth}{!}{%
\begin{tikzpicture}
\begin{axis}[
    bar width = 7pt,
    xbar=0,
    xlabel = speedup relative to the Direct Euler method \cite{boyce2017},
    xmode=log, 
    log basis x=10, 
    y=1.0cm,
    yticklabels = {\textsc{Spiral} (this paper), Buss \cite{buss2000accurate}, Fincham \cite{Fincham}, Johnson et al. \cite{johnson2008quaternion}, PFC4 \cite{PFC4}, Omelyan \cite{omelyan1998algorithm}, Velocity Verlet \cite{Verlet}, Runge-Kutta 4 \cite{kutta1901}}, 
    ytick = data,
    yticklabel style={align=center},
    axis y line*=none, 
    legend style={at={(0.5,1.05)}, anchor=south, legend cell align=left, legend columns=2, column sep=0.2cm},
    legend image code/.code={\draw [#1] (0cm,-0.1cm) rectangle (0.2cm,0.25cm); },
]
\addplot[fill=green!60!white, draw=black] coordinates { 
    (10.97, 0) 
    (10.25, 1) 
    (1.75, 2) 
    (0.55, 3) 
    (0.07, 4) 
    (4.1, 5) 
    (2.28, 6) 
};
\addplot[fill=blue!60!white, draw=black] coordinates { 
    (69.5, 0) 
    (13.6, 1) 
    (1.9, 2) 
    (0.57, 3) 
    (0.07, 4) 
    (15.03, 5) 
    (5.3, 6) 
};
\addplot[fill=orange!60!white, draw=black] coordinates { 
    (331, 0) 
    (12.4, 1) 
    (1.89, 2) 
    (0.57, 3) 
    (0.07, 4) 
    (56.2, 5) 
    (5.4, 6) 
};
\addplot[fill=red!70!white, draw=black] coordinates { 
    (1371, 0) 
    (13.2, 1) 
    (1.9, 2) 
    (0.58, 3) 
    (0.07, 4) 
    (175.0, 5) 
    (5.2, 6) 
};
\legend{error = $10^{-2}$, error = $10^{-3}$, error = $10^{-4}$, error = $10^{-5}$};
\end{axis}
\end{tikzpicture}}
\caption{Speedup of several algorithms relative to the Direct Euler method. The speedup is computed by measuring the Computer time needed to simulate 1\,s of the system specified in Fig. \ref{fig:analytical_sol_plot} with a predefined target error of $(\norm{q-q^\prime} + \norm{\vec{\omega}-\vec{\omega^\prime}})/2$ at the end of the simulation. The shown speedup is the time required by the Direct Euler scheme \cite{boyce2017} divided by the time required by each algorithm, and is shown for four values of the target errors: $10^{-2}$ (green), $10^{-3}$ (blue), $10^{-4}$ (orange) and $10^{-5}$ (red). The time step each algorithm requires to achieve the different target errors is provided in Tab. \ref{tab:dt-target}. The errors were computed by using Eq. \eqref{eq:error_metric}.
}
\label{fig:performance3}
\end{figure}

Although \textsc{Spiral} takes about $30\,\%$ more time to perform a single time step compared to the Direct Euler method \cite{boyce2017}, it ultimately allows for faster simulations. This is because \textsc{Spiral} requires fewer steps to achieve the same level of accuracy as other methods, as quantified in the previous Section. For a fair evaluation of the algorithms' efficiency, we determine the appropriate time step that achieves a given error by using the data from Fig. \ref{fig:dt_error}, with the bracketing algorithm from \textsc{Roots.jl} \cite{Roots.jl}. Measuring the total CPU time needed for each algorithm to integrate a particle's motion for a period of $1$\,\si{s} with a predefined total error, we can asses their relative speedups. Fig. \ref{fig:performance3}  shows the speedup of each algorithm as the time spent with Direct Euler \cite{boyce2017} divided by the time spent by that algorithm to reach the same target error after $1$\,\si{s}. The time step required for a list of specified target errors is reported in Tab. \ref{tab:dt-target}. Again, \textsc{Spiral} reveals much better than all other algorithms, around 10 times faster than Omelyan \cite{omelyan1998algorithm} and more than 100 times faster than Velocity Verlet \cite{Verlet}. Observe that all algorithms except Omelyan and \textsc{Spiral} are of the same order of accuracy than Direct Euler and, therefore, scale the same as Direct Euler when decreasing the target error. 


\begin{table}[htbp]
    \centering
    \caption{The time step $\Delta t$ required to achieve the average target error $\left(\norm{q-q^\prime|}+\norm{\vec{\omega}-\vec{\omega^\prime}}\right)/2$ after $1$\,\si{s} of simulation time for the problem described in Sec. \ref{sec:analytical_sol}. The error was computed using Eq. \eqref{eq:error_metric}.
    }
    \begin{tabular}{p{4cm}|c c c c}
        \hline
        \multicolumn{1}{p{3cm}}{ } & \multicolumn{4}{c}{target error} \\
        \cline{1-5}
        algorithm & $10^{-5}$ & $10^{-4}$ & $10^{-3}$ & $10^{-2}$ \\
        \hline
        \textsc{Spiral} (this paper) & 3.044e-03 & 8.372e-03 & 2.329e-02 & 6.167e-02 \\
        Omelyan \cite{omelyan1998algorithm} & 2.943e-04 & 9.319e-04 & 2.952e-03 & 9.400e-03 \\
        Buss \cite{buss2000accurate} & 2.650e-05 & 2.663e-04 & 2.890e-03 & 5.529e-02 \\
        Velocity Verlet \cite{Verlet} & 6.789e-06 & 6.852e-05 & 6.779e-04 & 3.527e-03 \\
        Fincham \cite{Fincham} & 3.069e-06 & 3.071e-05 & 3.091e-04 & 3.295e-03 \\
        Direct Euler \cite{boyce2017} & 1.324e-06 & 1.325e-05 & 1.333e-04 & 1.421e-03 \\
        Johnson et al. \cite{johnson2008quaternion} & 1.801e-06 & 1.801e-05 & 1.804e-04 & 1.824e-03 \\
        PFC4 \cite{PFC4,ostanin2023rigid} & 1.662e-07 & 1.662e-06 & 1.663e-05 & 1.671e-04 \\
        \hline
    \end{tabular}
    \label{tab:dt-target}
\end{table}

\clearpage

\section{Energy conservation of a many-particle system}\label{sec:ManyParticleTest}

\begin{figure}[htbp]
\centering
\includegraphics[width=0.9\textwidth]{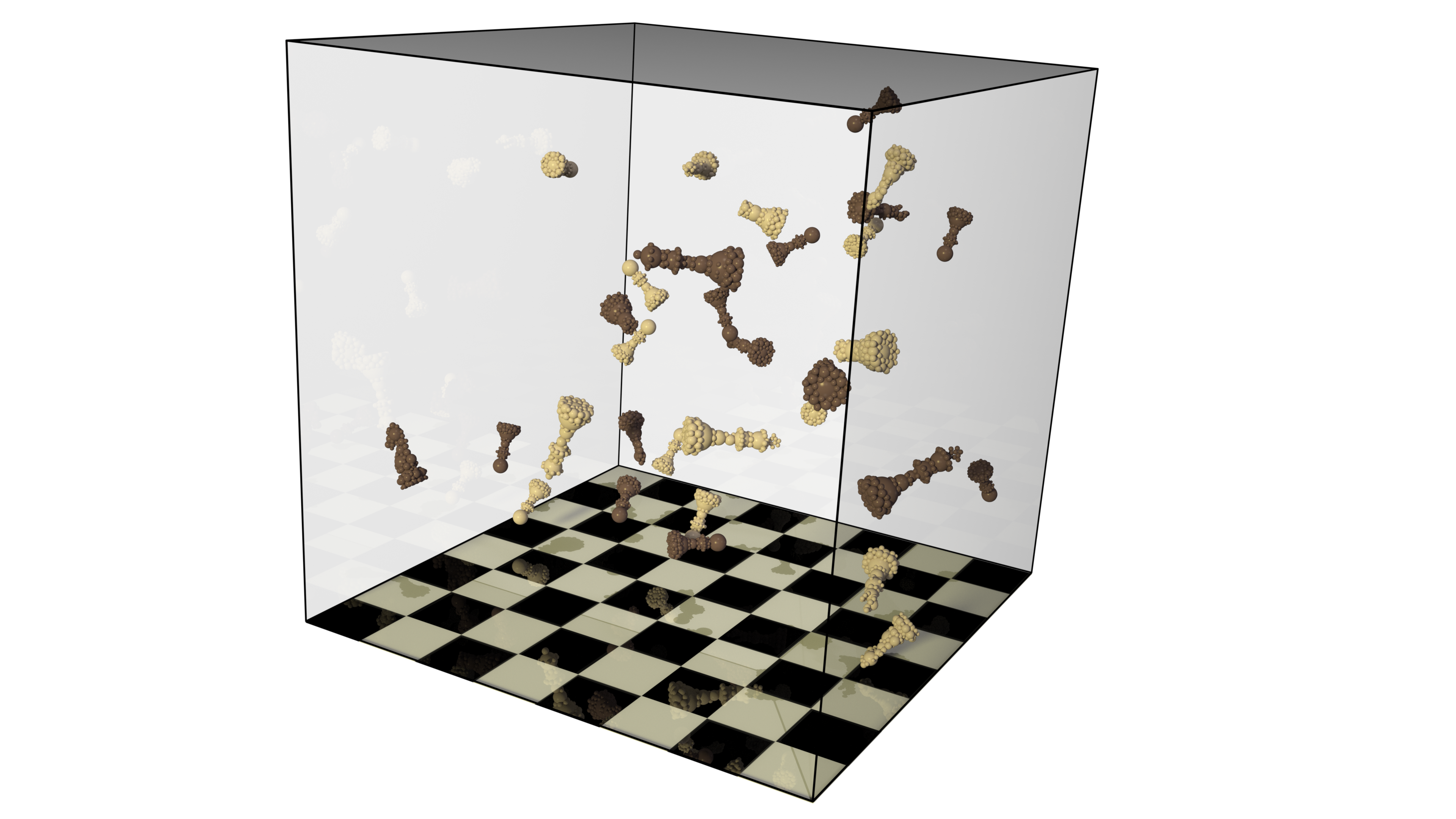}
\caption{Snapshot of a simulation of non-spherical particles (chess pieces modeled as multi-spheres) bouncing in a box. The elastic Hertz contact force describes particle-particle and particle-wall collisions.}
    \label{fig:ChessPieces}
\end{figure}

To evaluate the performance of \textsc{Spiral} in a many-body environment, we simulated non-spherical particles in a box. The contacts between the particles and the box, as well as between the particles themselves, are described by the conservative Hertz law, ensuring energy conservation. Our focus was on studying energy conservation in a dynamic system of chess pieces, as illustrated in Fig. \ref{fig:ChessPieces}.

The sizes of the chess pieces follow FIDE recommendations (king: $9.5$\,\si{cm}, queen $8.5$\,\si{cm}, bishop $7$\,\si{cm}, knight $6$\,\si{cm}, rook $5.5$\,\si{cm}, pawn $5$\,\si{cm}). Material density of the pieces is $3200$\,\si{kg/m^3}, the Young modulus, $60$\,\si{GPa}, and the Poisson ratio,\,$0.25$. We used \textsc{Yade} \cite{YADE} as a simulation framework. Initially, the positions of the objects were random, such that the particles were not in contact. Their initial velocity was $1$\,\si{m/s} in a random direction. Their initial angular velocity was $\pi$\,\si{rad/s}, also in a random direction.

\begin{figure}[H]
    \centering
    \begin{tikzpicture}
    \begin{axis}[
        xlabel={$\Delta t$ [s]},
        ylabel={\% energy change},
        legend style={at={(1.05,0.5)}, anchor=west},
        xmode=log,
        ymode=log,
        yminorticks=false,
        xminorticks=false,
        every axis plot/.append style={very thick},
        mark options={
            scale=1.5,               
            very thick,                 
        }
    ]
        \addplot [only marks, mark=diamond, blue] table [x=Time_Fincham, y=Fincham, col sep=comma] {Energy.csv};
        \addlegendentry{Fincham \cite{Fincham}}


        \addplot [only marks, mark=triangle, green!60!black] table [x=Time_Omelyan, y=Omelyan, col sep=comma] {Energy.csv};
        \addlegendentry{Omelyan \cite{omelyan1998algorithm}}

        \addplot [only marks, mark=pentagon, red] table [x=Time_SPIRAL, y=SPIRAL, col sep=comma] {Energy.csv};
        \addlegendentry{\textsc{Spiral} (this paper)}

        \draw[dashed, blue!70!black] ({axis cs:1.9e-6,0}|-{rel axis cs:0,1}) -- ({axis cs:1.9e-6,0}|-{rel axis cs:0,0}) node [above] {Rayleigh time step};
    \end{axis}
    \end{tikzpicture}
      \caption{Relative deviation of the kinetic energy \% energy from the initial value due to a simulation time of 1,5\,s as a function of the integration time step width. The results are averaged about 250 random initial conditions. The graph shows several integration schemes. A dashed line indicates the Rayleigh time step. 
    }
    \label{fig:energy_stability}
\end{figure}
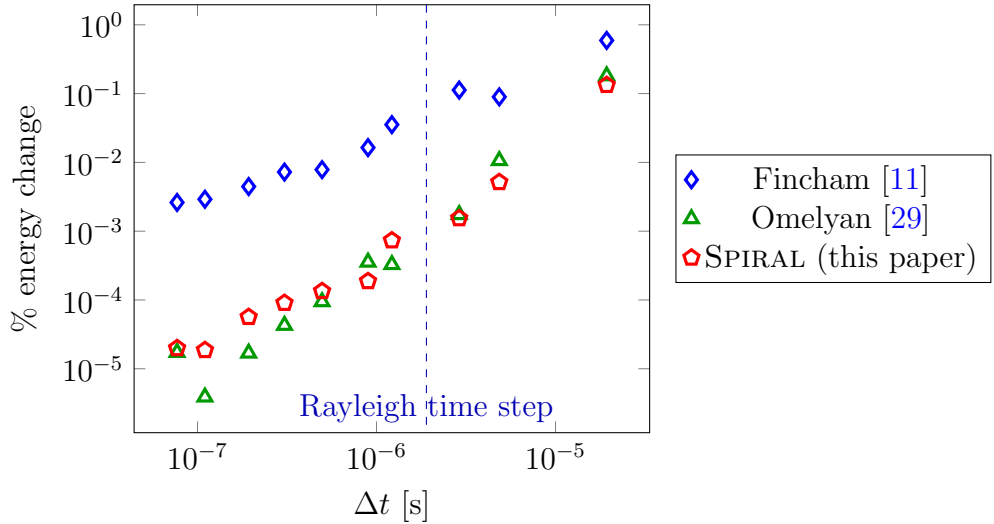

Fig. \ref{fig:energy_stability} shows the deviation from the initial kinetic energy as a function of the time step size for three integration algorithms: Fincham, Omelyan and \textsc{Spiral}. To facilitate comparison, the dashed line in the figure represents the Rayleigh time step \cite{Thornton2013}. The Rayleigh time step, $\Delta t_c$ is the time taken for a shear wave to propagate through a solid particle. It is common to use $\Delta t_c$ as a criterion to select a time step $\Delta t \sim 0.1\Delta t_c$ to $0.3\Delta t_c$ when looking for simulation stability.

The graph displays the relative change of energy in a total simulation time of $1.5$\,\si{s}. We see that \textsc{Spiral} conserves energy on the same level as the integration scheme by Omelyan \cite{omelyan1998algorithm}, which is known to be very stable.

\section{Conclusions}

We introduced \textsc{Spiral}, a highly accurate and efficient third-order integration algorithm for rigid body rotational motion. The algorithm displays excellent numerical stability. We provide both leapfrog and non-leapfrog variants for easy adaptation to simulation frameworks. Our algorithm features remarkable accuracy and stability while incurring minimal performance penalties compared to other algorithms used in popular DEM software; see Tab. \ref{tab:algorithms}. By construction, our algorithm preserves the norm of the quaternion, thus removing the need for regular quaternion normalization in each time step. The stability and preservation of the quaternion norm also enable the use of larger time steps, which improves simulation performance while achieving the same level of accuracy as other integration schemes. In addition, \textsc{Spiral} employs a modified third-order Strong-Stability Preserving Runge-Kutta (SSPKR3) scheme to update the angular velocity, which computes the torques only once per time step. When combined, those integration schemes give \textsc{Spiral} an outstanding performance: 10 times faster than Omelyan \cite{omelyan1998algorithm} and 100 times faster than Velocity Verlet \cite{Verlet} to reach the same target error.

We tested the algorithm stability in two scenarios: first, by looking at the accumulation of errors against time for a single rotor with analytical solution (the same we used for studying accuracy) and, second, by testing for energy conservation on a realistic DEM scenario of non-spherical particles interacting with a conservative Hertz law. In the first test, our algorithm was superior to all other algorithms tested and, in the second one, it was on par with the Omelyan algorithm \cite{omelyan1998algorithm}, which is known to reach excellent energy conservation.

As part of this work, we integrated \textsc{Spiral} algorithm into two major DEM simulation codes. Now, \textsc{Spiral} is an integral part of the \textsc{Yade} \cite{YADE} and \textsc{MercuryDPM} \cite{MercuryDPM} distributions, ensuring that our algorithm's benefits are accessible to a broad scientific community.

\section{Acknowledgements}
CAV thanks the Institute for Multiscale Simulation (MSS) at the Friedrich-Alexander-Universit\"at Erlangen-N\"urnberg for funding a visitor's grant.
 
\section{Data availability}

The \textsc{Spiral} algorithm was implemented in two major DEM simulation systems \textsc{Yade} and \textsc{MercuryDPM}. The implementation of \textsc{Spiral} as tested in this paper and the corresponding test scripts are available at \url{https://github.com/cdelv/AlgorithmsForRotationalMotion}. 

\section{Conflict of interest declaration}

We declare we have no competing interests.

\bibliographystyle{ajl}
\bibliography{references}

\clearpage

\appendix

\section{Non-leapfrog variant of \textsc{Spiral}}\label{app:Non-leapfrog}

To derive a non-leapfrog variant of \textsc{Spiral}, we follow the steps in Sec. \ref{sec:AlgorithmDerivation}. With $a=t$ in Eq. \eqref{eq:TaylorOmega} we obtain 
\begin{equation}
        q(t + \Delta t) = q(t)\,\exp\left[\frac{\Delta t}{2}\omega(t) + \frac{\Delta t^2}{4}\dot{\omega}(t) + \mathcal{O}(\Delta t^3)\right]\,,
\end{equation}
which is the non-leapfrog version of Eq. \eqref{eq:leap}. Using Eq. \eqref{eq:QuaternionPureRotation} we obtain
\begin{equation}
 q(t+\Delta t) = q(t)\left(\cos{\theta_2} + \frac{\vec{\omega}\left(t\right)}{\left|\vec{\omega}\left(t\right)\right|}\sin{\theta_2} \right)\left(\cos{\theta_3} + \frac{\dot{\vec{\omega}}\left(t\right)}{\left|\dot{\vec{\omega}}\left(t\right)\right|}\sin{\theta_3} \right)
\end{equation}
with
\begin{equation}
\theta_2 = \frac{\Delta t}{2}\left|\vec{\omega}\left(t\right)\right|\,,\qquad
\theta_3 = \frac{\Delta t^2}{4}\left|\dot{\vec{\omega}}\left(t\right)\right|\,.
\end{equation}
The non-leapfrog variant does not need a backward step for initialization. At each time step, $q(t)$ is updated before $\vec{\omega}(t)$. This order is necessary as the torque is required to compute $\dot{\vec{\omega}}$ with Eq. \eqref{eq:ang_vel_derivative}, which is only known at time $t$. Figs.  \ref{fig:dt_error2} and \ref{fig:qw_erorr_time2} 
compare the leap-frog and non-leapfrog versions of \textsc{Spiral} with the scheme by Omelyan \cite{omelyan1998algorithm}, which was the second best performing in Sec. \ref{sec:analytical_sol}, after \textsc{Spiral}. We observe that both versions of \textsc{Spiral} perform similarly in accuracy and error accumulation.
\begin{figure}[htbp]
    \centering
    \resizebox{\textwidth}{!}{%
    \begin{tikzpicture}
        \begin{groupplot}[
            group style={
                group size=2 by 1, 
                horizontal sep=2cm, 
            },
            axis x line=bottom,
            axis y line=left,
            xlabel=$\Delta t$  {[\si{s}]}, 
            no markers,
            every axis plot/.append style={very thick},
            xmode=log,
            ymode=log,
            yminorticks=false,
            xminorticks=false
        ]
        
        \nextgroupplot[ylabel=relative error: $\norm{q - q^\prime}$]
        \addplot [color=blue, densely dashed] table [x=dt, y=Omelyanq, col sep=comma] {Latex_dt_error.csv}; \label{plot:Omelyanq2}
        \addplot [color=black] table [x=dt, y=SPIRALq, col sep=comma] {Latex_dt_error.csv}; \label{plot:SPIRALq2}
        \addplot [color=red, densely dashed] table [x=dt, y=SPIRALLq, col sep=comma] {Latex_dt_error.csv}; \label{plot:SPIRALLq2}
        
        \nextgroupplot[ylabel=relative error:~ $\norm{\vec{\omega} - \vec{\omega}^\prime}$]
        \addplot [color=blue, densely dashed] table [x=dt, y=Omelyanw, col sep=comma] {Latex_dt_error.csv}; \label{plot:Omelyanw2}
        \addplot [color=red, densely dashed] table [x=dt, y=SPIRALLw, col sep=comma] {Latex_dt_error.csv}; \label{plot:SPIRALLw2}
        \addplot [color=black] table [x=dt, y=SPIRALw, col sep=comma] {Latex_dt_error.csv}; \label{plot:SPIRALw2}
        \end{groupplot}
        
        \node[align=center,anchor=north] at ($(group c1r1.south)!0.5!(group c2r1.south) + (0,-1.2cm)$) {%
         \begin{tabular}{@{}lll@{}}
                \ref{plot:Omelyanq2} Omelyan \cite{omelyan1998algorithm} & \ref{plot:SPIRALq2} \textsc{Spiral} (non-LeapFrog) & \ref{plot:SPIRALLq2} \textsc{Spiral} \\
        \end{tabular}};
    \end{tikzpicture}}
    \caption{Comparison of \textsc{Spiral} (leapfrog and non-leap frog variants) with the leapfrog scheme by Omelyan \cite{omelyan1998algorithm}: (left) relative error of the predicted quaternion with respect to the semi-analytical solution as a function of the time step, (right) relative error of the angular velocity with respect to the analytical solution as a function of the time step. The error was computed according to Eq. \eqref{eq:error_metric}.}
\label{fig:dt_error2}
\end{figure}
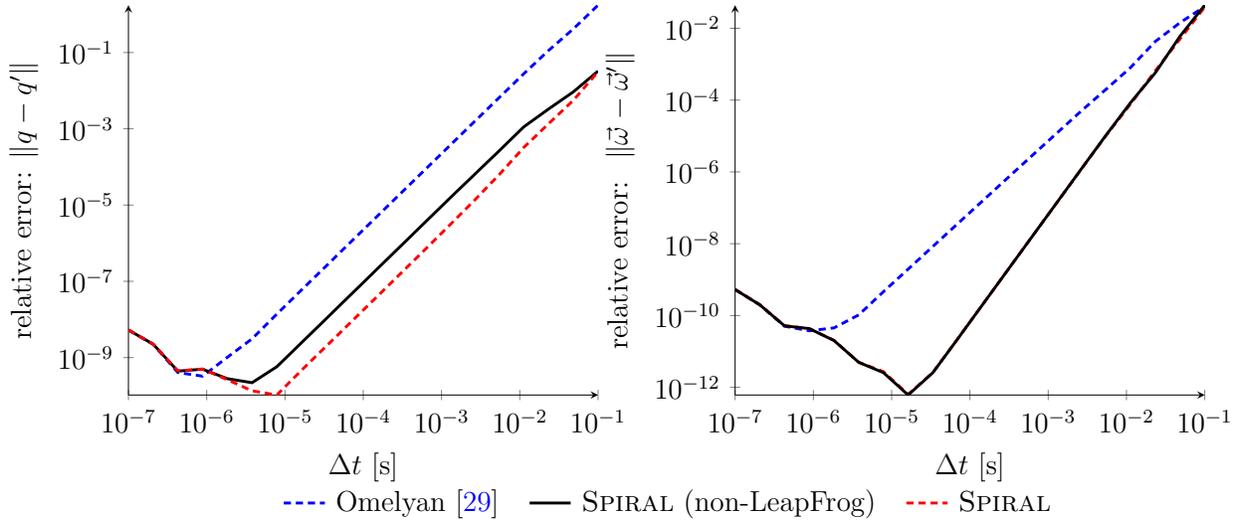
\begin{figure}[htbp]
    \centering
    \resizebox{\textwidth}{!}{%
    \begin{tikzpicture}
        \begin{groupplot}[
            group style={
                group size=2 by 1, 
                horizontal sep=2cm, 
            },
            axis x line=bottom,
            axis y line=left,
            xlabel=total simulation time $t$ {[\si{s}]}, 
            no markers,
            every axis plot/.append style={very thick},
            ymode=log,
            yminorticks=false,
            xminorticks=false
        ]
        
        \nextgroupplot[ylabel=relative error: $\norm{q - q^\prime}$]
        \addplot [color=blue, densely dashed] table [x=Omelyan_Time, y=Omelyan_Error_q, col sep=comma] {qw_error_time.csv}; \label{plot:Omelyanqt2}
        \addplot [color=red, densely dashed] table [x=SPIRAL_L_Time, y=SPIRAL_L_Error_q, col sep=comma] {qw_error_time.csv}; \label{plot:SPIRALLqt2}
        \addplot [color=black] table [x=SPIRAL_L_Time, y=SPIRAL_Error_q, col sep=comma] {qw_error_time.csv}; \label{plot:SPIRALqt2}

        \nextgroupplot[ylabel=relative error: $\norm{\vec{\omega} - \vec{\omega^\prime}}$]
        \addplot [color=blue, densely dashed] table [x=Omelyan_Time, y=Omelyan_Error_w, col sep=comma] {qw_error_time.csv}; \label{plot:Omelyanwt2}
        \addplot [color=black] table [x=SPIRAL_L_Time, y=SPIRAL_Error_w, col sep=comma] {qw_error_time.csv}; \label{plot:SPIRALwt2}
        \addplot [color=red, densely dashed] table [x=SPIRAL_L_Time, y=SPIRAL_L_Error_w, col sep=comma] {qw_error_time.csv}; \label{plot:SPIRALLwt2}
        \end{groupplot}
        
        \node[align=center,anchor=north] at ($(group c1r1.south)!0.5!(group c2r1.south) + (0,-1.2cm)$) {%
         \begin{tabular}{@{}lll@{}}
                \ref{plot:Omelyanqt2} Omelyan \cite{omelyan1998algorithm} & \ref{plot:SPIRALqt2} \textsc{Spiral} (non-LeapFrog) & \ref{plot:SPIRALLqt2} \textsc{Spiral} \\
        \end{tabular}};
    \end{tikzpicture}}
    \caption{Comparison of \textsc{Spiral} (leapfrog and non-leap frog variants) with the leapfrog scheme by Omelyan \cite{omelyan1998algorithm}; evolution of the error. The time step is $\Delta t = 10^{-5}$\,\si{s}. (left) Relative error of the predicted quaternion with respect to the semi-analytical solution as a function of total simulation time. (right) Relative error of the angular velocity with respect to the analytical solution as a function of the total simulation time. The error was computed according to Eq. \eqref{eq:error_metric}.}
    \label{fig:qw_erorr_time2}
\end{figure}
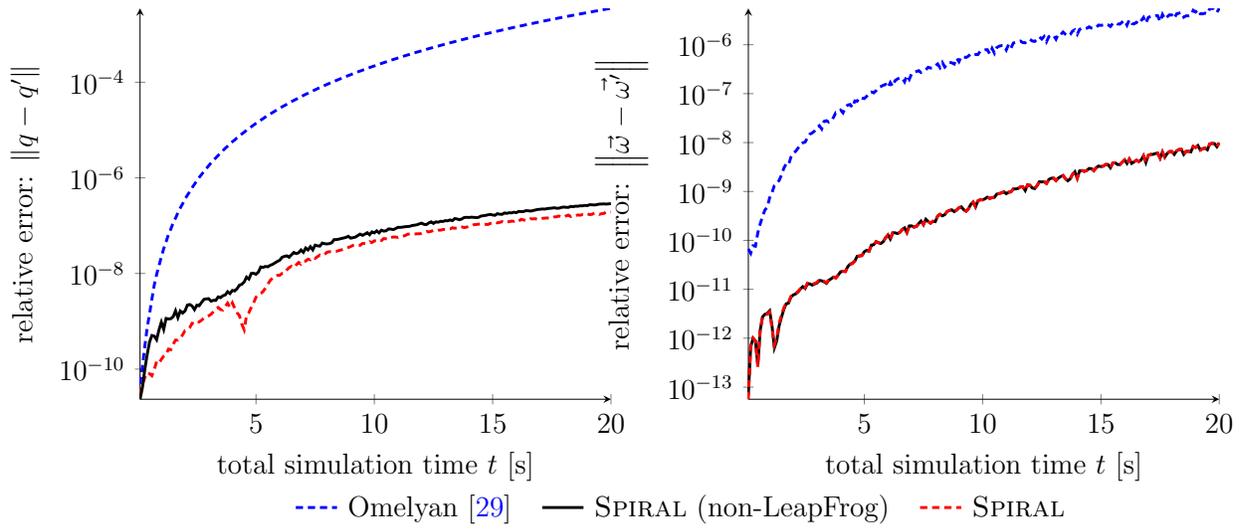
\clearpage

\section{Rotation Algorithms}\label{sec:Rotation_algorithms}
This Appendix provides details of the algorithms listed in \autoref{tab:algorithms} and used for comparison in Secs. \ref{sec:analytical_sol}, \ref{sec:Performance}, \ref{sec:ManyParticleTest}, and \ref{app:Non-leapfrog}.

To prevent any confusion regarding the reference frames used in the algorithms, we establish the following notation: Quantities described in the principal axis reference frame of the rotating object appear like in the rest of this paper (the angular velocity $\vec{\omega}$, for instance). Variables in the lab frame have a superscript, $\vec{\omega}^\text{lab}$. 
Furthermore, in the principal axis frame, the inertia tensor is represented as $\vec{I}$, and it is treated mathematically as a vector, because it is a diagonal matrix. In the lab frame, the inertia tensor is denoted as $J$ and is a 3x3 symmetrical matrix. The quaternion representing the orientation is the same for both reference frames.

The angular velocity derivative, as described in Eq. \ref{eq:ang_vel_derivative}, is formulated in the principal axis reference frame. This derivative is a function of the inertia tensor, the angular velocity, and the torque. Similarly, the quaternion derivative in Eq. \ref{eq:Dotq} is also formulated in the principal axis frame. To calculate the derivative in the lab frame, you can use the equation $\dot{q} = \frac{1}{2}q\omega$. It is important to recognize that this equation, along with Eq. \ref{eq:Dotq}, requires the angular velocity to be converted into an imaginary quaternion. Alternatively, the quaternion derivative can be expressed by using matrix-vector multiplication. Consequently, you can represent $\dot{q}$ as a function of the quaternion and the angular velocity vector \cite{quaternion_pdf}. For clarity, in the algorithm description, we show how to evaluate each derivative with brackets (for example, $\dot{\vec{\omega}}\left[\vec{\omega}, \vec{M}\right]$).

\subsection{Direct Euler}
The second-order Direct Euler algorithm \cite{boyce2017} updates the angular velocities and the quaternions through
\begin{equation}
\begin{split}
\vec{\omega}(t + \Delta t) &= \vec{\omega}(t) + \Delta t\,\dot{\vec{\omega}}\left[\vec{\omega}(t),\vec{M}(t)\right] \\
q(t + \Delta t) &= q(t) + \Delta t\,\dot{q}\left[q(t), \vec{\omega}(t + \Delta t)\right]\,.           
\end{split}
\end{equation}
This algorithm requires normalization of the quaternions in each time step. All quantities are in the principal axis frame.

\subsection{Velocity Verlet}
The second-order Velocity Verlet algorithm \cite{Verlet} updates the angular velocities and the quaternions through
\begin{equation}
\begin{split}
\vec{\omega}\left(t + \frac{\Delta t}{2}\right) &= \vec{\omega}(t) + \frac{\Delta t}{2}\,\dot{\vec{\omega}}\left[\vec{\omega}(t),\vec{M}(t)\right] \\
q(t + \Delta t) &= q(t) + \Delta t\, \dot{q}\left[q(t), \vec{\omega}(t + \Delta t)\right]\,.
\end{split}
\end{equation}
Compute the torques, $\vec{M}(t + \Delta t)$, and then
\begin{equation}
 \vec{\omega}(t + \Delta t) = \vec{\omega}\left(t + \frac{\Delta t}{2}\right) + \frac{\Delta t}{2}\dot{\vec{\omega}}\left[\vec{\omega}\left(t + \frac{\Delta t}{2}\right),\vec{M}(t + \Delta t)\right]\,.
 \end{equation}
This algorithm requires normalization of the quaternions in each time step. All quantities are in the principal axis frame.

\subsection{Fincham Leapfrog}

The second-order Fincham Leapfrog algorithm \cite{Fincham} uses angular momenta instead of angular velocities. It includes a mid-step interpolation of the quaternion to increase the stability. The update from $t$ to $t+\Delta t$ is performed in two steps. First, the angular momenta are updated as in Leapfrog schemes,
\begin{equation}
\vec{L}^\text{lab}\left(t + \frac{\Delta t}{2}\right) = \vec{L}^\text{lab}\left(t - \frac{\Delta t}{2}\right) + \Delta t\, \vec{M}^\text{lab}(t)\,.
\end{equation}
In this algorithm, the torque and angular momentum are in the lab reference frame. Then, we update the quaternion with a predictor-corrector scheme. Here, we need the derivative of the quaternion at different angular velocities,
\begin{equation}
\begin{split}
&q_a = q + \frac{\Delta t}{2}\,\dot{q}\left[q(t), \vec{\omega}(t)\right] \, \\
&q(t + \Delta t) = q + \Delta t\,\dot{q}\left[q_a, \vec{\omega}(t)\right].
    \end{split}
\end{equation}
The angular velocities in the rotating body-fixed frame can be calculated as 
\begin{equation}
\begin{split}
&\vec \omega(t) =\hat{A}\left[\vec{L}\left(t - \frac{\Delta t}{2}\right) + \frac{\Delta t}{2}\,\vec{ M}(t)\right] \, \\
&\vec \omega\left(t + \frac{\Delta t}{2}\right) = \hat{A}\, \vec{L}\left(t + \frac{\Delta t}{2}\right)\,,
\end{split}
\end{equation}
where $\hat{A}$ is the rotation matrix to translate from lab-fixed to body-fixed coordinates. The matrix can be calculated from the quaternion components. This algorithm requires normalization of the quaternion.

\subsection{Buss Algorithm}
Unlike the above-described algorithms, the second-order Buss algorithm \cite{buss2000accurate} updates the inertia tensor and uses the quantities in the lab frame. Before the update, we compute
\begin{equation}
\begin{split}
J(t) &= \hat{A}^{-1}(t)\, \vec{I} \, \\
\vec{\omega}^\text{lab}(t) &= J^{-1}(t)\,\vec{L}^\text{lab}(t)\, \\
\dot{\vec{\omega}}^\text{lab}(t) &= J^{-1}(t)\left[\vec{M}^\text{lab}(t) - \vec{\omega}^\text{lab}(t) \times \vec{L}^\text{lab}(t)\right]\,.
\end{split}
\end{equation}
Here, $\hat{A}$ is the rotation matrix for transforming lab-fixed  into body-fixed coordinates, and $J$ is the inertia tensor matrix in the lab frame, calculated from the inertia tensor in the principal axis frame $\vec{I}$. The rotation matrix can be computed from the quaternion components. 

The update proceeds in two steps. First, let us compute
\begin{equation}
\begin{split}
\bar{\vec{\omega}} &= \vec{\omega}^\text{lab}(t) + \frac{\Delta t}{2}\,\dot{\vec{\omega}}^\text{lab}(t) + \frac{\Delta t^2}{12}\left[\dot{\vec{\omega}}^\text{lab}(t)\times \vec{\omega}^\text{lab}(t)\right] \, \\
\theta &= \left|\bar{\vec{\omega}}\right|\,\Delta t
\end{split}
\end{equation}
and then
\begin{equation}
\begin{split}
q(t + \Delta t) &= \left[\cos{\frac{\theta}{2}} + \frac{\vec{\bar{\omega}}}{|\Vec{\bar{\omega}}|}\sin{\frac{\theta}{2}}\right]\,q(t) \, \\[3mm]
\vec{L}^\text{lab}(t + \Delta t) &= \vec{L}^\text{lab}(t) + \Delta t\,\vec{M}^\text{lab}(t)\,.
\end{split}
\end{equation}

The version of the Buss algorithm presented here is the same as the one in PFC7 documentation \cite{PFC}. Buss' original version \cite{buss2000accurate} uses rotation matrices instead of quaternions.

\subsection{Omelyan Algorithm}
Omelyan's third-order Advance Leapfrog algorithm \cite{omelyan1998algorithm} proceeds as follows:
\begin{equation}
\label{eq:Omelyan98_algorithm}
\begin{split}
\vec{\omega}\left(t+\frac{\Delta t}{2}\right) &= \vec{\omega}\left(t-\frac{\Delta t}{2}\right) + \Delta t\,\dot{\vec{\omega}}\left[\vec{\omega}(t),\vec{M}(t)\right] \, \\
q(t+\Delta t) &= \frac{\left(1-\frac{\Delta t^2}{16}\left| \vec{\omega}\left(t+\frac{\Delta t}{2}\right)\right|^2\right)\mathbbm{1} + \Delta t\, \dot{q}\left[q(t), \vec{\omega}(t + \frac{\Delta t}{2})\right]}{1 + \frac{\Delta t^2}{16} \left|\vec{\omega}\left(t+\frac{\Delta t}{2}\right)\right|^2}\,q(t)\, ,
\end{split}
\end{equation}
where $\mathbbm{1}$ is the identity quaternion. Eq. \eqref{eq:Omelyan98_algorithm} uses the derivative of the angular velocity at time $t$. Comparison with Eq. \eqref{eq:ang_vel_derivative} reveals that the computation of $\dot{\vec{\omega}}(t)$ would require 
the angular velocity at time $t$, which is yet unknown. To solve this problem, Omelyan proposed the approximation
\begin{equation}
\label{eq:Omelyan_aprox}
\omega_\alpha(t)\,\omega_\beta(t) \approx \frac{1}{2}\left[\omega_\alpha\left(t-\frac{\Delta t}{2}\right)\omega_\beta\left(t-\frac{\Delta t}{2}\right) + \omega_\alpha\left(t+\frac{\Delta t}{2}\right)\omega_\beta\left(t+\frac{\Delta t}{2}\right)\right].
\end{equation}
This approximation turns Eqs. \eqref{eq:Omelyan98_algorithm} into a system of nonlinear equations that can be efficiently solved by iteration, with the initial guess
\begin{equation}
    \vec{\omega}^{(0)}(t) = \vec{\omega}\left(t-\frac{\Delta t}{2}\right)\,.
\end{equation}
Experience shows that \-- for a sufficiently small time step \-- about three iterations are enough to achieve convergence. In this paper, we always used three iterations.

By construction, this algorithm preserves the norm of the quaternions. As a result, renormalization shouldn't be necessary; however, we found that the algorithm accumulates significant numerical errors over time and can become unstable. We suggest sporadically normalizing the quaternion when using this algorithm. 

\subsection{Johnson (2008) Algorithm}
Johnson's algorithm \cite{johnson2008quaternion} is based on a Runge-Kutta-4 scheme for the update of the quaternion:
\begin{equation}
  q(t + \Delta t) = q(t) + \frac{\Delta t}{6}\left(k_1 + 2k_2 + 2k_3 +k_4\right)\,,
\end{equation}
with
\begin{equation}
\begin{split}
k_1 &= \dot{q}\left[q(t),\; \vec \omega(t)\right]\, \\
k_2 &= \dot{q}\left[q(t) + \Delta t\,\frac{k_1}{2},\; \vec{\omega}(t)\right] \, \\
k_3 &= \dot{q}\left[q(t) + \Delta t\,\frac{k_2}{2},\; \vec{\omega}(t)\right] \, \\
k_4 &= \dot{q}\left[q(t) + \Delta t\,k_3,\; \vec{\omega}(t)\right]\,.
\end{split}
\end{equation}
Next, the angular momentum is updated as
\begin{equation}
    \vec L^\text{lab}(t + \Delta t) = \vec L^\text{lab}(t) + \Delta t  \vec M^\text{lab}(t)\,.
\end{equation}

Note that Johnson suggests to use a second-order predictor-corrector algorithm for evolving angular momentum \cite{johnson2008quaternion}. This would require two force calculations and two quaternion updates per time step; nevertheless, for the tests in Sec. \ref{sec:analytical_sol} the algorithm wouldn't have benefited from that predictor-corrector scheme, because the torque is given and constant. This algorithm requires the quaternion to be normalized. 

\subsection{PFC4/ MercuryDPM}
The algorithm used by MercuryDPM \cite{MercuryDPM} for non-spherical particles is described in the PFC4 manual \cite{PFC4} and \cite{ostanin2023rigid}. It uses rotation matrices to represent the particle orientation. We are not aware of the algorithm's name or its original reference. The current version of PFC \cite{PFC} uses either the Buss \cite{buss2000accurate} or the Johnson algorithm \cite{johnson2008quaternion}. In PFC4 all the quantities are in the lab reference frame. 

The algorithm proceeds as follows:
First, the angular velocity and momentum are iteratively updated until convergence as
\begin{equation}
\begin{split}
\vec{L}_n^\text{lab} &= J\,\vec{\omega}_n^\text{lab} \, \\
\vec{\omega}_{n+1}^\text{lab} &= \vec{\omega}_0^\text{lab} + \Delta t\,J^{-1}\left(\vec{M}^\text{lab} - \vec{\omega}_{n}^\text{lab}\times \vec{L}_{n}^\text{lab}\right)\,.
\end{split}
\end{equation}
Then, the orientation is updated as
\begin{equation}
    \hat{A}(t + \Delta t) = \hat{A}(t) + \Delta t\,\dot{\hat{A}}(t)\,,
\end{equation}
where $\hat{A}$ is the rotation matrix representing the particle's orientation. Its derivative reads
\begin{equation}
    \dot{\hat{A}} = \varepsilon_{ijk}\,\omega_j^\text{lab}\, \hat{A}_{mk}\,,
\end{equation}
where $\varepsilon_{ijk}$ is the Levi-Civita symbol. Einstein's summation notation applies. In this paper, we used three iterations for convergence, as in MercuryDPM.
\clearpage
 
\section{A special solution of Euler's equations of motion}
\label{app:AnalyticalSolution}
\subsection{The case of constant driving, $M_z\ne 0$}

We consider a rigid body with principal moments of inertia $I_x\ne I_y = I_z$, driven by a constant torque $\vec{M} = \left\{M_x,0,0\right\}$ with $M_x\ne 0$. Under these assumptions, Euler's rigid body equations of motion in the principal axis frame, Eq. \eqref{eq:ang_vel_derivative}, reduces to
\begin{eqnarray}
\label{eq:rigid_body_simplified-x}
I_x\,\dot{\omega}_x &=& M_x \,\\
\label{eq:rigid_body_simplified-y}
I_y\,\dot{\omega}_y &=& \omega_z\,\omega_x\,\left(I_z - I_x\right) \,\\
\label{eq:rigid_body_simplified-z}
I_z\,\dot{\omega}_z &=& \omega_x\,\omega_y\,\left(I_x - I_y\right)\,.
\end{eqnarray}
The solution of Eq. \eqref{eq:rigid_body_simplified-x} is trivial,
\begin{equation}\label{eq:omega_x}
    \omega_x(t) = \omega_x^0 + \frac{M_x}{I_x}\,t\,,
\end{equation}
with $\omega_x^0\equiv \omega_x(0)$.
To solve Eqs. (\ref{eq:rigid_body_simplified-y}, \ref{eq:rigid_body_simplified-z}), take the time derivative of Eq. \eqref{eq:rigid_body_simplified-y}, 
\begin{equation}
\label{eq:TPapp1}
    \frac{I_y}{I_z - I_x}\,\ddot{\omega}_y = \dot{\omega}_z \omega_x + \omega_z \dot{\omega}_x\,,
\end{equation}
and substitute $\dot{\omega_z}$ obtained from this equation into Eq. \ref{eq:rigid_body_simplified-z}, to obtain
\begin{equation}
\label{eq:TPapp2}
\frac{I_z}{I_x - I_y}\left( \frac{I_y}{I_z - I_x}\,\frac{1}{\omega_x}\,\ddot{\omega}_y - \frac{\dot{\omega}_x}{\omega_x}\,\omega_z \right) = \omega_x\,\omega_y \,.
\end{equation}
We also substitute $\omega_z$ as obtained from Eq. \eqref{eq:rigid_body_simplified-y} into Eq. \eqref{eq:TPapp2}, to obtain
\begin{equation}
\label{eq:TPapp3}
\frac{I_z}{I_x - I_y}\left( \frac{I_y}{I_z - I_x}\,\frac{1}{\omega_x}\ddot{\omega}_y - \frac{\dot{\omega}_x}{\omega_x}\frac{I_y}{I_z - I_x}\frac{1}{\omega_x}\dot{\omega}_y \right) = \omega_x\omega_y \,.
\end{equation}
Equation \eqref{eq:TPapp3} is true if $\omega_x \ne 0$. It can be written conveniently as 
\begin{equation}
\label{eq:differential_eq}
\ddot{\omega}_y - \frac{\dot{\omega}_x}{\omega_x}\,\dot{\omega}_y  = B\,\omega_x^2\,\omega_y \,
\end{equation}
with
\begin{equation}
B\equiv \frac{(I_x - I_y)(I_z - I_x)}{I_zI_y}\,.
\end{equation}
According to our initial assumption, $I_x\ne I_y = I_z$, we define $I_*\equiv I_y = I_z$ and, thus,
\begin{equation}
\label{eq:B-lt-0}
B= - \left(\frac{I_x - I_*}{I_*}\right)^2 < 0\,.
\end{equation}

Since $\omega_x(t)$ is an explicit function of time (Eq. \eqref{eq:omega_x}), we can use $\omega_x$ as a time scale and replace derivatives $\text{d}/\text{d}t$ by $\text{d}/\text{d}\omega_x$. We obtain
\begin{equation}
\label{eq:TPd-d-omega_x}
    \begin{split}
        \frac{\text{d}\omega_y}{\text{d} t} &= \frac{\text{d} \omega_y}{\text{d}\omega_x}\,\frac{\text{d}\omega_x}{\text{d}t}
        =\frac{M_x}{I_x}\,\frac{\text{d}\omega_y}{\text{d}\omega_x} \, \\
        \frac{\text{d}^2\omega_y}{\text{d} t^2} &= \left(\frac{M_x}{I_x}\right)^2\, \frac{\text{d}^2 \omega_y}{\text{d}\omega_x^2}\,.
    \end{split}
\end{equation}
Equation \eqref{eq:differential_eq} reads, then,
\begin{equation}
\label{eq:differential_eq_TP}
    \frac{\text{d}^2\omega_y}{\text{d}\omega_x^2} 
    - \frac{1}{\omega_x}\,\frac{\text{d}\omega_y}{\text{d}\omega_x}
    =\left(\frac{I_x}{M_x}\right)^2 B\, \omega_x^2\,\omega_y\,,
\end{equation}
provided $M_x\ne 0$. 
Substituting $\tau\equiv \omega_x^2$ yields
\begin{equation}
\label{eq:TPstar4}
\begin{split}
\frac{\text{d} \omega_y}{\text{d}\omega_x} &= \frac{\text{d} \omega_y}{\text{d}\tau} \,\frac{\text{d} \tau}{\text{d}\omega_x}
        = 2\,\omega_x \,\frac{\text{d} \omega_y}{\text{d}\tau} \,\\
\frac{\text{d}^2 \omega_y}{\text{d}\omega_x^2} &= 2 \,\frac{\text{d} \omega_y}{\text{d}\tau} + 2\,\omega_x \frac{\text{d}}{\text{d}\omega_x}\,\frac{\text{d} \omega_y}{\text{d}\tau}
= 2\,\frac{\text{d} \omega_y}{\text{d}\tau} + 4\tau \,\frac{\text{d}^2 \omega_y}{\text{d}\tau^2}\, ,        
\end{split}
\end{equation}
and Eq. \eqref{eq:differential_eq_TP} becomes
\begin{equation}
\label{eq:TP14}
\frac{\text{d}^2 \omega_y}{\text{d}\tau^2} = \left(\frac{I_x}{2 M_x}\right)^2 B\,\omega_y\,.
\end{equation}

The last expression is the equation of motion for a harmonic oscillator, which we recast as
\begin{equation}
\frac{\text{d}^2 \omega_y}{\text{d}\tau^2} = - \Omega^2 \,\omega_y\, ,
\end{equation}
with 
\begin{equation}
\label{eq:Omega-def}
\Omega^2\equiv -\left(\frac{I_x}{2 M_x}\right)^2 B\,.
\end{equation}
(Recall that $B<0$, see Eq. \eqref{eq:B-lt-0}).
Its solution reads 
\begin{equation}
\label{eq:omega_y_tau}
\omega_y(\tau) = k_1 \cos \left(\Omega\tau\right) + k_2 \sin\left(\Omega\tau\right)\, ,
\end{equation} 
whose coefficients $k_1$ and $k_2$ shall be determined from the initial conditions. We derive a corresponding equation from Eq. \eqref{eq:rigid_body_simplified-y} by using Eqs. (\ref{eq:TPd-d-omega_x},\ref{eq:TPstar4},\ref{eq:TP14}), obtaining
\begin{equation}
\omega_z=\frac{I_y}{I_z-I_x} \,\frac{1}{\omega_x} \, \frac{\text{d}\omega_y}{\text{d}t} 
=\frac{I_y}{I_z-I_x} \,\frac{1}{\omega_x} \, \frac{M_x}{I_x}\, \frac{\text{d}\omega_y}{\text{d}\omega_x}
=\frac{I_y}{I_z-I_x} \,2\,\frac{M_x}{I_x}\, \frac{\text{d}\omega_y}{\text{d}\tau}
=\eta  \, \frac{\text{d}\omega_y}{\text{d}\tau}\,,
\end{equation}
with 
\begin{equation}
\label{eq:eta-def}
    \eta\equiv \frac{I_y}{I_z-I_x} \,\frac{2\,M_x}{I_x}\,.
\end{equation}
Thus, from Eq. \eqref{eq:omega_y_tau} we obtain
\begin{equation}
\label{eq:omega_z_tau}
\omega_z(\tau) = \eta\Omega \left(k_2 \cos\right(\Omega\tau\left) - k_1 \sin\right(\Omega\tau\left)\right)\,. 
\end{equation}

The set of Eqs. (\ref{eq:omega_x}, \ref{eq:omega_y_tau}, \ref{eq:omega_z_tau}) form the complete solution of Euler's Equations (\ref{eq:rigid_body_simplified-x},\ref{eq:rigid_body_simplified-y},\ref{eq:rigid_body_simplified-z}).
For $t=0$, we find $\tau=\left(\omega_x^0\right)^2$, according to its definition. With this result on hand, we obtain $\omega_y^0\equiv \omega_y(t=0)$ and $\omega_z^0\equiv \omega_z(t=0)$ from Eqs. (\ref{eq:omega_y_tau},\ref{eq:omega_z_tau}) as
\begin{equation}
\label{eq:omega_yz_0}
\begin{split}
\omega_y^0 &= k_1\,\cos\left(\Omega\left(\omega_x^0\right)^2\right) + k_2\,\sin\left(\Omega\left(\omega_x^0\right)^2\right) \,\\    
\omega_z^0 &= \eta\Omega\left[k_2\,\cos\left(\Omega\left(\omega_x^0\right)^2\right) - k_1\,\sin\left(\Omega\left(\omega_x^0\right)^2\right)\right]\,.
\end{split}
\end{equation}
By solving the linear set of Equations \eqref{eq:omega_yz_0} we determine the parameters $k_1$ and $k_2$ from the initial velocities $\left(\omega_x^0, \omega_y^0, \omega_z^0\right)$. With the definitions of $\Omega$ and $\eta$ (Eqs. \eqref{eq:Omega-def} and \eqref{eq:eta-def}) we get
\begin{equation}
    \begin{split}
        k_{1} &= \omega_{y}^0 \cos\left(\Omega \left(\omega_x^0\right)^2\right) - \frac{\omega_{z}^0}{\Omega \eta}\sin\left(\Omega \left(\omega_x^0\right)^2\right)\, \\
        k_{2} &= \omega_{y}^0 \sin\left(\Omega \left(\omega_x^0\right)^2\right) + \frac{\omega_{z}^0}{\Omega \eta}\cos\left(\Omega \left(\omega_x^0\right)^2\right),
    \end{split}
\end{equation}
which concludes the analytical solution of Eqs. (\ref{eq:rigid_body_simplified-x},\ref{eq:rigid_body_simplified-y},\ref{eq:rigid_body_simplified-z}).

\subsection{The undriven rotator, $M_x=0$}
For this special case, we can see from \eqref{eq:rigid_body_simplified-x} that
\begin{equation}
\label{eq:omegax-case}
\omega_x(t)=\omega_x^0\,.     
\end{equation}
By inserting $\dot{\omega}_z$ from Eq. \eqref{eq:rigid_body_simplified-z} into the time derivative of Eq. \eqref{eq:rigid_body_simplified-y}, we obtain
\begin{equation}
\ddot{\omega_y}=\frac{I_z-I_x}{I_y} \omega_x^0 \dot{\omega}_z
=\left(\omega_x^0\right)^2 \frac{I_z-I_x}{I_y}\frac{I_x-I_y}{I_z} \,\omega_y \, .
\end{equation}
This is again a harmonic oscillator equation,
\begin{equation}
\label{eq:omegaY-case}
\ddot{\omega_y} = -\Gamma^2\,\omega_y\,,
\end{equation}
with 
$\Gamma>0$ for $I_x \ne I_y = I_z$. Using its formal solution for $\omega_y(t)$, we can solve the equation 
\begin{equation}
\label{eq:omegaZ-case}
    \omega_z(t)= \frac{I_y}{I_z - I_x}\frac{1}{\omega_x^0}\dot{\omega_y} = \zeta \dot{\omega_y} \,
\end{equation}
and, thereafter, the set (\ref{eq:omegax-case},\ref{eq:omegaY-case}, \ref{eq:omegaZ-case}) can be solved in the same way as shown in the preceding subsection.


\end{document}